\documentclass[final,twoside]{IEEEtran}
\usepackage{standalone}
\usepackage{booktabs,longtable}

\usepackage{epstopdf}

\usepackage{cite}
\usepackage{amsmath}
\usepackage{amssymb}

\usepackage{xcolor}
\usepackage{lipsum}

\usepackage{siunitx}
\usepackage{pgfplots}
\pgfplotsset{compat=newest}
\usepackage{pgfplotstable}
\usepgfplotslibrary{groupplots}
\usepgfplotslibrary{external} 
\usetikzlibrary{patterns}
\usepgfplotslibrary{external, groupplots} 
\usetikzlibrary{patterns}

\usetikzlibrary{spy,backgrounds}

\DeclareGraphicsExtensions{.pdf,.eps,.jpg,.png}

\usepackage{hyperref}
\hypersetup{
    bookmarks=true,         
    unicode=false,          
    pdftoolbar=true,        
    pdfmenubar=true,        
    pdffitwindow=false,     
    pdfstartview={FitH},    
    pdftitle={Physical bounds for antenna radiation efficiency},    
    pdfauthor={Shahpari, Thiel},     
    pdfsubject={},   
    pdfcreator={Shahpari},   
    pdfproducer={Shahpari}, 
    pdfkeywords={}, 
    pdfnewwindow=true,      
    colorlinks=true,        
    linkcolor=blue,         
    citecolor=red,          
    filecolor=magenta,      
    urlcolor=cyan           
}
\RequirePackage[normalem]{ulem} 
\RequirePackage{color}\definecolor{RED}{rgb}{1,0,0}\definecolor{BLUE}{rgb}{0,0,1} 

\begin{document}

\graphicspath{{Figs4Paper/}}
\title{Fundamental limitations for antenna radiation efficiency}

\author{Morteza Shahpari \IEEEmembership{Member, IEEE}
	and David V. Thiel \IEEEmembership{Life Senior Member, IEEE}
	\thanks{This work was partly supported by Australian research council discovery project (DP130102098).}
	\thanks{M. Shahpari and D. V. Thiel are with school of engineering, Griffith University, Gold Coast campus, QLD, Australia,
		Tel: +61 7 5552 8459,
		morteza.shahpari@ieee.org,d.thiel@griffith.edu.au}}

\markboth{IEEE Trans. Antennas Propag. Vol. 66, NO. X, 2018}{Shahpari and Thiel: Fundamental limitations for antenna radiation efficiency}
\maketitle

\begin{abstract}
Small volume, finite conductivity and high frequencies are major imperatives in the design of communications infrastructure. 
The radiation efficiency $\eta_r$ impacts on the optimal gain, quality factor, and bandwidth.
The current efficiency limit applies to structures confined to a radian sphere $ka$ (where $k$ is the wave number, $a$ is the radius). 
Here, we present new fundamental limits to $\eta_r$ for arbitrary antenna shapes based on $k^2S$ where $S$ is the conductor surface area. 
For a dipole with an electrical length of $10^{-5}$ our result is  {two} orders of magnitude closer to the analytical solution when compared with previous bounds on the efficiency. 
The improved bound on $\eta_r$ is more accurate, more general, and easier to calculate than other limits. 
The efficiency of an antenna cannot be larger than the case where the surface of the antenna is ‘peeled’ off and assembled into a planar sheet with area $S$, and a uniform current is excited along the surface of this sheet.
\end{abstract}

\begin{IEEEkeywords}
	Antenna efficiency, upper bound, efficiency, fundamental limit, conductivity, skin depth.
\end{IEEEkeywords}	
\section{INTRODUCTION}
\label{introduction}
\IEEEPARstart{I}{N a world} relying more and more on wireless communications, antenna efficiency is of central importance in predicting radio communications reliability.
Based on IEEE Standard for Definitions of Terms for Antennas~\cite{AntDef_2014_IEEESTD}, radiation efficiency $\eta_r$ is ``the ratio of the total power radiated by an antenna to the net power accepted by the antenna from the connected transmitter.''
Unless exactly specified, the term efficiency means radiation efficiency throughout the paper.
New fabrication techniques, new materials and smaller antennas are of significant interest to reduce e-waste and to make fabrication easier. A low efficiency antenna has reduced gain and so the communications range is reduced. In portable mobile platforms, most battery power is related to radiation. If the efficiency is increased, the battery life increases. This is also of great interest by communications specialists in the trade-off between fabrication costs, antenna size and antenna efficiency. As we show in this paper, the current methods used to predict the maximum possible efficiency based on the radian sphere highly overestimate performance when the conductivity is finite or relatively low for non-spherical objects. We have developed a new approach to the calculation of maximum antenna efficiency so new technologies can be assessed reliably and compared to the maximum possible efficiency determined from the improved fundamental limits.

Unlike the quality factor $Q$ which is extensively studied, the fundamental limits on the maximum radiation efficiency of an antenna has not been extensively studied~\cite{Volakis_b_2010,Gustafsson_2015_SpringerChapter,Shahpari_2015_Thesis}. 
As illustrated in \cite{Shahpari_2015_TAP}, the radiation efficiency directly impacts on various antenna parameters. 
Therefore, a robust limit on $\eta_{r}$ also complements the limitations on bandwidth \cite{Harrington_1960_JRNBS,Hansen_1981_PIEEE}, gain \cite{Geyi_2003_TAP_PhysLim,Pigeon_2014_IJMWT}, $Q$ factor \cite{Wheeler_1947_ProcIEEE,Chu_1948_JAP,Collin_1964_TAP,Fante_1969_TAP,McLean_1996_TAP,Thal_2006_TAP,Thal_2012_TAP,Kim_2016_TAP,Jonsson_2015_RSPA,Hansen_2011_APM,Yaghjian_2013_PIER,Yaghjian_2005_TAP}, and gain over \(Q\) ratio~\cite{Geyi_2003_TAP_PhysLim,Jonsson_2015_RSPA,Gustafsson_2007_RSPA,Gustafsson_2009_TAP}. 
Optimization algorithms employed to achieve highly efficient antennas~\cite{Lewis_bc_2009} can be quantified by comparing results with the fundamental physical bound. 
Physical bounds also provide simple rules to check the feasibility of a specific product requirement with the given material conductivity and dimension.

Harrington~\cite{Harrington_1960_JRNBS} initiated studies on the limitations imposed by a lossy medium on the antenna efficiency. 
Arbabi and Safavi-Naeini~\cite{Arbabi_2012_TAP} approached the problem from another point of view. 
They used a spherical wave expansion in a lossy medium to find the dissipated power, and consequently $\eta_{r}$. 
Fujita and Shirai~\cite{Fujita_2015_IEICE} added a non-radiating term to study the effect of the antenna shape. 
They concluded that the spherical shape is an optimum shape which has a
potential to maximize the antenna efficiency. 
A similar approach to maximize $\eta_{r}$ was proposed in~\cite{Karlsson_2013_PIER} by seeking an optimum current distribution for spherical shapes.
Pfeiffer~\cite{Pfeiffer_2017_TAP} and Thal
\cite{Thal_2018_TAP} incorporated the effect of metallic loss in Thal equivalent circuits~\cite{Thal_1978_TAP,Thal_2006_TAP} to find the maximum antenna efficiency.
The results are also extended for spherical metallic shell antennas.

The common points in the previous works~\cite{Harrington_1960_JRNBS,Arbabi_2012_TAP,Fujita_2015_IEICE} are that they assume the lossy medium still holds the good conductor condition. 
Also, these works only focus on spherical antennas. 
To find the limiting values~\cite{Harrington_1960_JRNBS,Arbabi_2012_TAP,Fujita_2015_IEICE}, spherical Bessel and Hankel functions were integrated using the properties of the Bessel functions. 
However, their final result is still cumbersome to find by an engineering calculator.
On the other hand, the derivations from the equivalent circuit~\cite{Pfeiffer_2017_TAP} arrives to a simple closed form formula.

In this paper, we derive a fundamental limit on the antenna efficiency. 
Unlike most of previous works, our calculations provide closed form solutions for the limiting radiation efficiency values.
Our limit can also be used for all shapes including non-spherical geometries. Therefore, our new physical bound can be used to predict the limiting performance of spheroidal, cylindrical, and even planar structures of finite thickness.
A similar approach is used to find the maximum efficiency of infinitely thin structures.

Organization of the paper is as follows: wave equation and propagation of the wave in a lossy media is briefly discussed in section~\ref{Sec_LossyMedia}. 
In section~\ref{Sec_CalcMaxEfficiency}, the maximum possible efficiency is derived with few approximations on dissipated and radiated power of a general antenna. 
A similar approach is followed in section~\ref{Sec_2DBound} to find efficiency of thin structure. 
Section~\ref{Sec_Results} illustrates the usefulness of the proposed fundamental limitations with examples of frequency or conductivity variations.
Electrical area \(k^{2}S\) was also introduced in subsection~\ref{Subsec_k2S} as an alternative for \(ka\) to scale antennas of arbitrary shapes. 
Variations of the maximum efficiency with frequency and electrical length $ka$ are also reported in the part~\ref{Subsec_CompareWOthers} where we also compare with previous bounds on the efficiency. 
Finally, we provide a direct comparison of the efficiency of the optimized planar structures~\cite{Gustafsson_2013_APS} with the proposed planar bounds in this paper.
\section{Propagation of the Wave in the lossy media}
\label{Sec_LossyMedia}

An arbitrary object with permittivity \(\epsilon\), permeability \(\mu\), and conductivity \(\sigma\) is assumed to occupy the volume \(V\) with the surface boundary \(S\). 
A time convention of $\mathrm{e}^{\mathrm{j}\omega t}$ is assumed.
Propagation of EM wave inside the object should satisfy the wave equation \(\nabla^{2}\boldsymbol{E}-\gamma^{2}\boldsymbol{E} = 0\) where \(\gamma = \alpha + \mathrm{j}\beta = \left( {-\omega}^{2}\mu\epsilon + \mathrm{j}\omega\mu\sigma \right)^{0.5}\).
For good conductors with \(\sigma \gg \omega\epsilon\), we can approximate real and imaginary parts of \(\gamma\) as \(\alpha \approx \beta \approx \left(\pi f \mu \sigma \right)^{0.5}\).

\begin{figure}
\centering
\begin{tabular}{cc}
\includegraphics{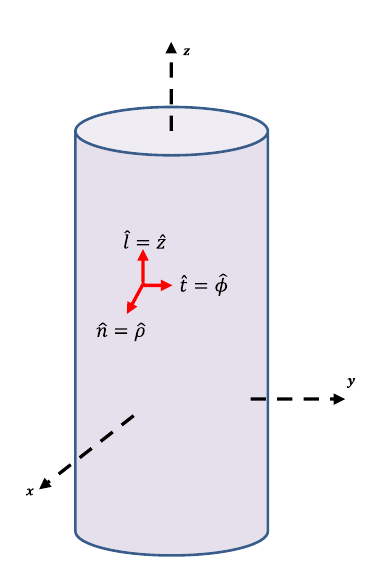}
&
\includegraphics[width=0.45\linewidth]{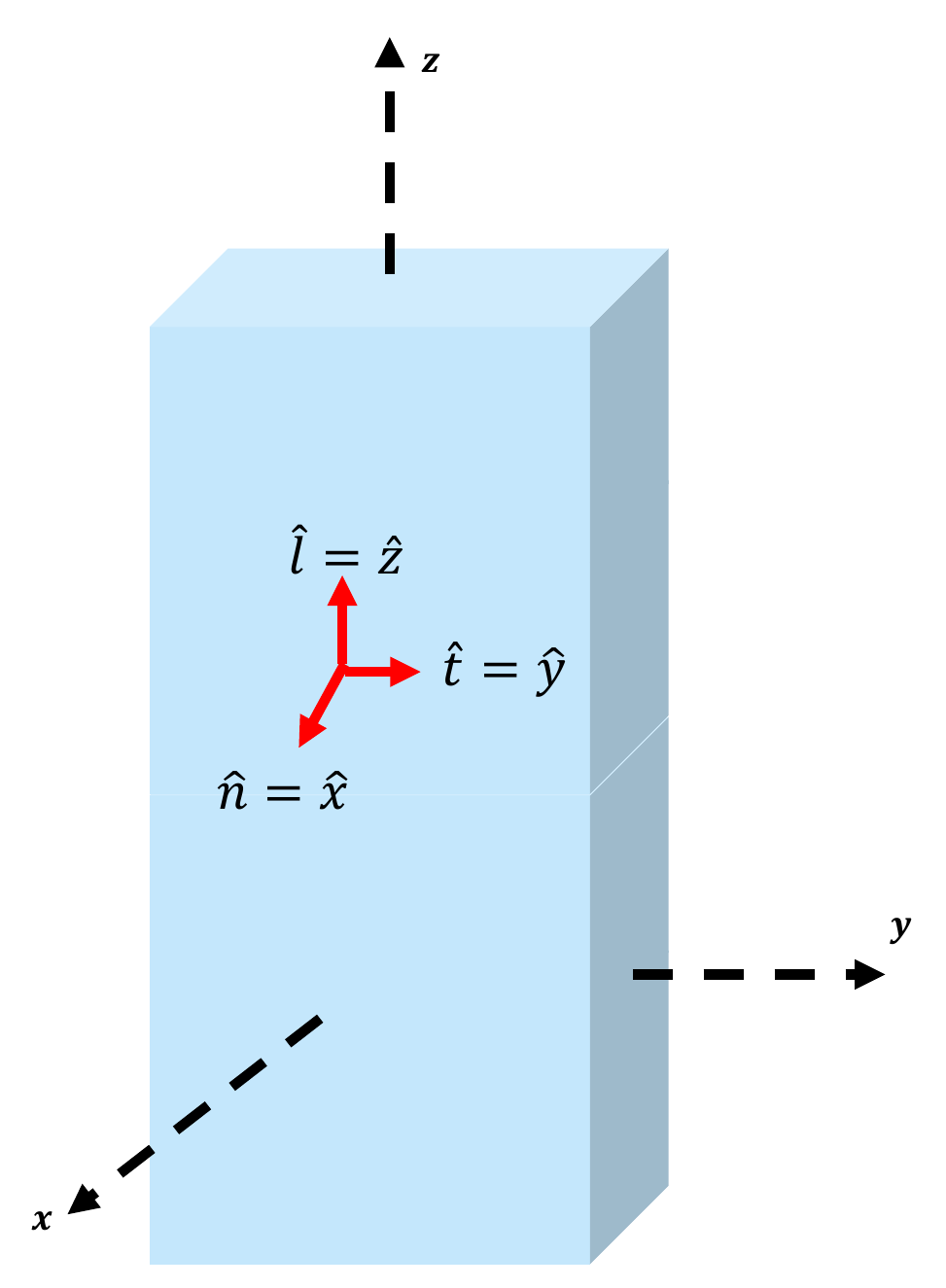}
\\
(a) & (b)
\end{tabular}
\caption{\textbf{The generalised coordinate system.} Our derivation of the new efficiency bounds uses surface current. While the antenna shape can be arbitrary two elemental shapes are shown for $\hat{\boldsymbol{n}}$, $\hat{\boldsymbol{t}}$, and $\hat{\boldsymbol{l}}$ definitions. (a) cylindrical geometry with ($n_0=r$), and (b) rectangular geometry with ($n_0=t$).}
\label{Fig_ArbGeom}
\end{figure}

Without losing generality, we consider a coordinate system constructed by the unit normal vector \(\hat{\boldsymbol{n}}\) and tangential vectors of  \(\hat{\boldsymbol{t}}\), and \(\hat{\boldsymbol{l}}\) where \(\hat{\boldsymbol{n}} \times \hat{\boldsymbol{t}} = \hat{\boldsymbol{l}}\)
(see Fig.\ref{Fig_ArbGeom}). 
We also assume that an arbitrary current \(\boldsymbol{J}\) (which satisfies Maxwell's equations) flows through the object and has values \(J_{s}\) on the surface \(S\) of the conducting object.
It should be noted that \(J_{s}\) has dimensions of \si{\ampere\per\meter\squared}, as it shows the values of the volume current on the boundaries of the medium. 
Due to the skin-effect phenomena, we can show that the current inside the volume \(V\) decays exponentially towards the centre of the object
\begin{align}
\left| \boldsymbol{J}\left( t,l,n \right) \right| = J_{s}\left( t,l \right)\mathrm{e}^{- \alpha\left( n_{0} - n \right)},
\label{Eq_JinMedium}
\end{align}
where \(n\) is the coordinate orthogonal to the object cross section, and \(n_{0}\) is the value of \(n\) on the surface \(S\). 
For instance, \(n_{0}\) can be considered as the radius of a cylinder and the thickness of the strip for cylindrical and planar structures,
respectively (see Fig.~\ref{Fig_ArbGeom}). 
The skin depth assumption in \eqref{Eq_JinMedium} is sought to be valid for frequencies up to far infrared region~\cite{Maier_b_2007}.

\section{Upper Bound on Efficiency of an Arbitrary Shaped Metallic Antenna}
\label{Sec_CalcMaxEfficiency}

\subsection{Dissipated Power}
\label{Subsec_dissipated-power}

One can find the power dissipated in the lossy material by using the Ohm law
\begin{align}
\label{Eq_Ploss_General}
P_{\text{loss}} &= \frac{1}{2\sigma}\iiint_V \left| \boldsymbol{J} \right|^{2} \mathrm{d}n\, \mathrm{d}t\, \mathrm{d}l,\\
& = \frac{1}{2\sigma} \iiint_{n=0}^{n=n_0} |J_s|^2\, \mathrm{e}^{2\alpha(n-n_0)} \mathrm{d}n\, \mathrm{d}t\, \mathrm{d}l.
\end{align}

Therefore, we find \(P_{\text{loss}}\)
\begin{align}
\label{Eq_Ploss}
P_{\text{loss}} = \frac{1}{4\sigma\alpha} \left[1-\mathrm{e}^{-2\alpha n_0}\right] \iint |J_s|^2 \, \mathrm{d}t\, \mathrm{d}l.
\end{align}

\subsection{Radiated Power}\label{radiated-power}

The radiated power can be calculated rigorously using the method introduced by Vandenbosch~\cite{Vandenbosch_2010_TAP}. 
This method is only based on the currents on the antenna (not farfield approximations of $\boldsymbol{E}$ and $\boldsymbol{H}$).
\begin{multline}
\label{Eq_Vandenbosh}
P_r = \frac{k}{8\pi \omega\mu_0} \int_{V_1} \int_{V_2} \left[k^2 \boldsymbol{J}(\boldsymbol{r}_1) \cdot \boldsymbol{J}^*(\boldsymbol{r}_2) \right.\\
- \nabla_1 \cdot \left.\boldsymbol{J}(\boldsymbol{r}_1) \nabla_2 \cdot \boldsymbol{J}^*(\boldsymbol{r}_2)\right] \frac{\sin(kR)}{kR} \mathrm{d}V_1\, \mathrm{d}V_2,
\end{multline}
where \(k = \omega\left( \mu_{0}\epsilon_{0} \right)^{0.5}\) is the wave number, and \(\boldsymbol{J}\) is the current flowing within the volume of the radiating device. 
The subscripts 1 and 2 indicate the first and the second of the double integration over the volume, and \(R\) is the distance between points 1 and 2 (\(R = \left| \boldsymbol{r}_{1} - \boldsymbol{r}_{2} \right|\)).
Characteristic impedance of the free space is also denoted by $\eta_0 =\sqrt{\mu_0/\epsilon_{0}}$.

For electrically small antennas \(kR \ll 1\), we use Taylor-McLaurin
expansion $\frac{\sin(kR)}{kR}\approx 1-\frac{(kR)^2}{6}+\frac{(kR)^4}{120}+\ldots$
By inserting only the first two terms in \eqref{Eq_Vandenbosh}, we have
\begin{align}
\label{Eq_TaylorExpandedPr}
P_r = \frac{\eta_0}{8\pi} &\int_{V_1} \int_{V_2} \left[k^2\boldsymbol{J}(\boldsymbol{r}_1) \cdot \boldsymbol{J}^*(\boldsymbol{r}_2)\right.\nonumber\\
&  +\frac{(kR)^2}{6} \nabla_1 \cdot \boldsymbol{J}(\boldsymbol{r}_1) \nabla_2 \cdot \boldsymbol{J}^*(\boldsymbol{r}_2) \big] \mathrm{d}V_1 \, \mathrm{d}V_2\nonumber\\
&  -\frac{\eta_0}{48\pi} \int_{V_1} \int_{V_2} (kR)^2 \boldsymbol{J}(\boldsymbol{r}_1) \cdot \boldsymbol{J}^*(\boldsymbol{r}_2) \, \mathrm{d}V_1 \, \mathrm{d}V_2\nonumber\\
&  -\frac{\eta_0}{8\pi} \int_{V_1} \int_{V_2} \nabla_1 \cdot \boldsymbol{J}(\boldsymbol{r}_1) \nabla_2 \cdot \boldsymbol{J}^*(\boldsymbol{r}_2) \mathrm{d}V_1 \, \mathrm{d}V_2.
\end{align}
The second integration is ignored since it is directly proportional to small term $(kR)^2$. 
The third integration in \eqref{Eq_TaylorExpandedPr} can be separated and rewritten as $\int_{V_1} \nabla_1 \cdot \boldsymbol{J}(\boldsymbol{r})_1 \, \mathrm{d}V_1 \int_{V_2} \nabla_2 \cdot \boldsymbol{J}^*(\boldsymbol{r}_2)  \, \mathrm{d}V_2$ which is always calculated as zero due to charge conservation law $\int_V \nabla\cdot\boldsymbol{J}(\boldsymbol{r}) \, \mathrm{d}V=0$.
One can use the following vector identity to simplify the first integration in \eqref{Eq_TaylorExpandedPr} (a proof is provided in the appendix):
\begin{multline}
\label{Eq_R2Div}
\int_{V_1} \int_{V_2} R^2 \nabla_1 \cdot\boldsymbol{J}(\boldsymbol{r}_1) \nabla_2 \cdot\boldsymbol{J}^*(\boldsymbol{r}_2) \, \mathrm{d}V_1 \, \mathrm{d}V_2 \\
=-2\int_{V_1} \int_{V_2}  \boldsymbol{J}(\boldsymbol{r}_1) \cdot \boldsymbol{J}^*(\boldsymbol{r}_2) \, \mathrm{d}V_1 \, \mathrm{d}V_2
\end{multline}
Therefore, we can find the radiated power as:
\begin{align}
P_r = &\frac{k^2 \eta_0}{12\pi} \int_{V_1} \boldsymbol{J}(\boldsymbol{r}_1) \, \mathrm{d}V_1 \, \cdot \int_{V_2} \boldsymbol{J}^*(\boldsymbol{r}_2) \, \mathrm{d}V_2\\
\label{Eq_Pr_inter}
= & \frac{k^2 \eta_0}{12\pi} \left|\int_V \boldsymbol{J} \, \mathrm{d}V \right|^2
\end{align}
A drawback of the approximations used above is that \eqref{Eq_Pr_inter} ignores radiation from loop like  currents.
However, since loops are far less efficient radiators than dipoles, this does not affect the upper bounds on the $P_{r_{max}}$ and efficiency.
By substituting~\eqref{Eq_JinMedium} in \eqref{Eq_Pr_inter}, we have:
\begin{align}
P_r &= \frac{k^2 \eta_0}{12\pi} \left[\iiint_{n=0}^{n_0} J_s \mathrm{e}^{\alpha(n-n_0)} \, \mathrm{d}n\, \mathrm{d}t\, \mathrm{d}l\right]^2\nonumber\\
& = \frac{k^2 \eta_0}{12\pi} \left[\iint J_s\, \mathrm{d}t\, \mathrm{d}l\right]^2 \left[\frac{1-\mathrm{e}^{-\alpha n_0}}{\alpha}\right]^2
\end{align}
If $f$ and $g$ are integrable complex functions, the Schwarz inequality allows:
\begin{align}
\left|\int f \, g^* \, \mathrm{d}x\right|^2\leq \int |f|^2 \, \mathrm{d}x \int |g|^2 \, \mathrm{d}x 
\end{align}
By assuming $f=J_s$ and $g=1$ as a constant, we can write:
\begin{align}
\label{Eq_UsedSchwarz}
\left|\iint_S J_s \, \mathrm{d}t \, \mathrm{d}l \right|^2 \leq S \iint_S |J_s|^2 \, \mathrm{d}t \, \mathrm{d}l
\end{align}

If \(J_{s}\) is constant then inequality \eqref{Eq_UsedSchwarz} becomes an equality. 
This is the case for Hertzian dipole antennas, while most of the small antennas have triangular distribution in practice.
Assuming triangular $(1-\frac{|z|}{l})$ and cosine $\cos(\frac{\pi z}{2l})$ distributions spanning from $-l$ to $l$, LHS of \eqref{Eq_UsedSchwarz} is $l^2$ and $\frac{16l^2}{\pi^2}$, respectively.
RHS of \eqref{Eq_UsedSchwarz} is found as $\frac{4l^2}{3}$ and $2l^2$.
Therefore, the approximation made in inequality \eqref{Eq_UsedSchwarz} results in almost 33\% and 23\% overestimation for triangular and cosine distributions, respectively. 
The overestimation is acceptable in the context of this contribution since we are looking for the highest radiated power from a structure. 
If the two sides of \eqref{Eq_UsedSchwarz} are far apart, then the synthesized current is not the optimum distribution.

Therefore, we can find the maximum radiated power \(P_{r_{\max}}\) from the structure:
\begin{align}
\label{Eq_Prmax3D}
P_{r_{max}} = \frac{\eta_0 k^2}{12\pi} \frac{\left[1-\mathrm{e}^{-\alpha n_0}\right]^2}{\alpha^2} S \iint_S |J_S|^2\, \mathrm{d}t\, \mathrm{d}l.
\end{align}

Radiation resistance found from \eqref{Eq_Prmax3D} exactly agrees with the radiation resistance of an infinitely small antenna with uniform distribution~\cite{Elliott_b_1982,Balanis_2005_antenna}.
It should be noted that $P_{r_{max}}$ from \eqref{Eq_Prmax3D} never goes to zero. 
Even if $\iint_S \boldsymbol{J} \cdot \mathrm{d}S=0$ (e.g. a small loop), we always have $|\boldsymbol{J}|^2 > 0$.
Since the radiation resistance of the small loops changes with $(ka)^4$, they are much less efficient than the electric dipoles $R_r \propto (ka)^2$.
Therefore, \eqref{Eq_Prmax3D} is the true maximum power radiated by any arrangement of $TM$ and $TE$ modes.

\subsection{Maximum Efficiency}\label{maximum-efficiency}

The radiation efficiency of an antenna is defined as: \(\eta_{r} = P_{r}/(P_{r} + P_{\text{loss}})\)~\cite{Galehdar_2007_AWPL}. 
Therefore, we can construct a bound on the radiation efficiency $\eta_{r}$ using \eqref{Eq_Ploss} and \eqref{Eq_Prmax3D}:
\begin{align}
\label{Eq_Eta_BoundGeneral}
\eta_{r_{max}} = 
\frac{
	\sigma \eta_0 k^2 S  \left[1-\mathrm{e}^{-\alpha n_0}\right]^2 
}{
	\sigma \eta_0 k^2 S \left[1-\mathrm{e}^{-\alpha n_0}\right]^2
	+
	3\pi \alpha [1-\mathrm{e}^{-2\alpha n_0}] }
\end{align}

For the majority of the antennas in the RF-microwave region, the skin depth is much smaller than the thickness of the conductor \(\delta \ll n_{0}\). 
Therefore, one can ignore \(\mathrm{e}^{- \alpha n_{0}}$ and $\mathrm{e}^{- 2\alpha n_{0}}\) terms in \eqref{Eq_Eta_BoundGeneral} as \(\alpha n_{0} \gg 1\)
\begin{align}
\label{Eq_Eta_Bound_HighAlphaN}
\eta_{r_{max}}	&= \frac{ \sigma \eta_0 k^2 S \delta}{ \sigma \eta_0 k^2 S \delta + 3\pi } = \left[1+ \frac{3\pi}{2} \frac{\delta}{kS}\right]^{-1}
\end{align}
In this paper, \eqref{Eq_Eta_BoundGeneral} is referred to as the general bound while \eqref{Eq_Eta_Bound_HighAlphaN} is quoted as the approximate limitation.

\section{Upper Bound on the Efficiency of 2D Antenna}
\label{Sec_2DBound}

A similar analysis is followed in this section to find maximum efficiency of infinitely thin antennas. 
Here, we assume the surface conductivity \(\sigma_{s}\) for the two-dimentional sheets of arbitrary currents. 
Therefore, the lost power can be rewritten from \eqref{Eq_Ploss_General} as:
\begin{align}
P_{\text{loss}} = \frac{1}{2\sigma_{s}}\iint\left| J_s \right|^{2} \, \mathrm{d}t \, \mathrm{d}l
\end{align}
One should note that the integration along the normal direction is omitted due to the zero thickness of the structure. 
A similar procedure is also repeated to find the maximum radiated power 
\begin{align}
P_r = \frac{\eta_0 k^2}{12\pi}\left[\iint \boldsymbol{J}_s \,\mathrm{d}t \, \mathrm{d}l \right]^2 = 
\frac{\eta_0 k^2}{12\pi} S \iint \left|\boldsymbol{J}_s \right|^2 \,\mathrm{d}t \, \mathrm{d}l
\end{align}
Therefore, the maximum efficiency is readily found as:
\begin{align}
\label{Eq_Bound2D}
\eta_{r_{max2D}} = \frac{\eta_{0}k^2 S\sigma_{s}}{\eta_{0}k^{2} S \sigma_{s} + 6\pi}
   = \left[1+ 3\pi \frac{\delta}{kS}\right]^{-1}
\end{align}

\section{Surface area $S$}                 
The surface area of the radiator $S$ plays a key role in the calculation of the maximum efficiency in this work which is clarified here.
In section~\ref{Sec_CalcMaxEfficiency}, $\iint_S |J_s|^2 \, \mathrm{d}t\, \mathrm{d}l$ runs over the sides of the objects $\hat{\boldsymbol{t}}$ and $\hat{\boldsymbol{l}}$, while the normal direction is taken care of through the skin depth effect.
For a single piece convex object like a prism, the area $S$ is the sum of the all of the exterior faces.

If the antenna consists consists of $N$ convex pieces (like Yagi-Uda antenna), then the area $S$ is the sum of the areas of different objects $S_j$, as long as the area $S$ does not exceed the area of the enclosing Chu sphere.
Therefore, $S$ is defined as:
\begin{align}
S= \min \left(4\pi a^2 , \sum_{j=0}^{N} S_j \right)
\end{align}

\section{Results}
\label{Sec_Results}

In this section, we elaborate on the implications of \eqref{Eq_Eta_BoundGeneral}, \eqref{Eq_Eta_Bound_HighAlphaN} and \eqref{Eq_Bound2D}.
We assumed \(\alpha n_{0}\gg 1\)  (thickness much larger than the skin depth) which leads to  $\exp(-\alpha n_0)\ll 1$ in finding \eqref{Eq_Eta_Bound_HighAlphaN} from \eqref{Eq_Eta_BoundGeneral}.
Colour in Fig 2 illustrates the values of \(10\log{(\alpha n_{0})}\) where the conductivity and frequency are varied for a material with thickness of \SI{600}{\micro\meter}. 
It is seen that the \(\alpha n_{0} \gg 1\) assumption is valid over a wide range of frequencies and conductivity for a relatively thin structure. 
As will be seen in the next subsections and graphs, \eqref{Eq_Eta_BoundGeneral} and \eqref{Eq_Eta_Bound_HighAlphaN} have close predictions while \(\alpha n_{0} \gg 1\). 
However, the approximate form diverges from the general formula when \(\alpha n_{0}\) lies in the range $\approx 1 - 5$ either by reducing frequency or the conductivity of the material.

\begin{figure}
	\centering
	\includegraphics[width=\linewidth]{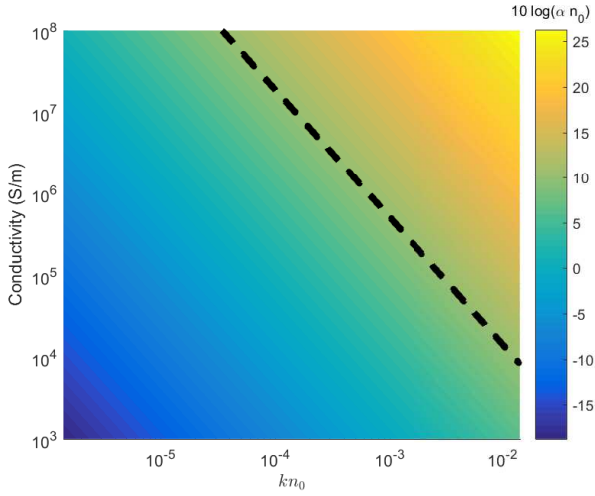}
	\caption{\textbf{Variation of $\alpha n_0$ with conductivity and normalized radius} for a cylindrical wire with radius $n_0=$\SI{0.6}{\milli\meter}. 
	Dashed line shows $\alpha n_0=$\SI{10}{\decibel} boundary. For copper wires with radius of \SI{0.6}{\milli\meter}, $\alpha n_0\gg 1$ is satisfied when $f\gg$\SI{3.6}{\mega\hertz}.
}
\label{Fig_alpha_n0}
\end{figure}

\subsection{Point of the maximum slope}
\label{Subsec_Point}

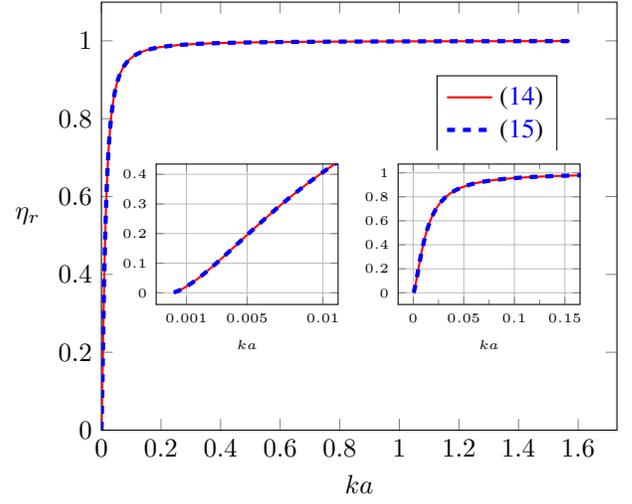
\begin{figure}
\centering
		\pgfplotsset{
		AntennaDesign/.style={
			mark repeat = 50},
		Eta_rLimit/.style={
			mark = no marks, red,thick},
		Eta_rLimit_HighAlphaN/.style={
			mark = no marks, blue,dashed,ultra thick},
		Eta_Arbabi/.style={
			mark= no mark,teal ,thick}
	}

\begin{tikzpicture}[every pin/.style={fill=white,pin edge=white}]

\begin{axis}[xlabel=$ka$,ylabel=$\eta_r$, y label style={rotate={-90}}, 
xmin=0, ymin=0]

	
\addplot [Eta_rLimit] table[x=ka,y=General] {Data_RadEff_Curve4Insets.dat};
	
\addplot [Eta_rLimit_HighAlphaN] table[x=ka,y expr=\thisrow{Eta_r_sigma_highAlphaN}] {Data_RadEff_Curve4Insets.dat};

\legend{\eqref{Eq_Eta_BoundGeneral},\eqref{Eq_Eta_Bound_HighAlphaN}}

\pgfplotsset{every axis legend/.append style={
	at={(0.65,.65)},
	anchor=south west}}

\coordinate (pt) at (axis cs:0.1,0.05);
\coordinate (pt2) at (axis cs:0.912,0.05);

\end{axis}

\node[pin=70:{%
    \begin{tikzpicture} [baseline,trim axis left,trim axis right]
    \begin{axis}[
        tiny,
	xlabel=$ka$,
      xmin=0,xmax=0.01,
      ymin=0,
      enlargelimits,
      xmajorgrids={true},
      xminorgrids={false},
      ymajorgrids={true},
      yminorgrids={false},
      minor x tick num=1,
      xtick={0.001,0.005,0.01},
      scaled ticks=false, tick label style={/pgf/number format/fixed},
      xticklabel style={/pgf/number format/fixed,
      	/pgf/number format/precision=3}
    ]
    
    \addplot +[Eta_rLimit] table[x=ka,y=General] {Data_RadEff_Curve4Insets.dat};
    \addplot +[Eta_rLimit_HighAlphaN] table[x=ka,y expr=\thisrow{Eta_r_sigma_highAlphaN}] {Data_RadEff_Curve4Insets.dat};

    \end{axis}
    \end{tikzpicture}%
}] at (pt) {};

\node[pin=70:{%
    \begin{tikzpicture}[baseline,trim axis left,trim axis right]
    \begin{axis}[
        tiny,
	xlabel=$ka$,
      xmin=0,xmax=0.15,
      ymin=0,
      enlargelimits,
        xmajorgrids={true},
      xminorgrids={false},
      ymajorgrids={true},
      yminorgrids={false},
      minor x tick num=1,
      scaled ticks=false, tick label style={/pgf/number format/fixed},
      xticklabel style={/pgf/number format/fixed,
      	/pgf/number format/precision=3}
        ]
    \addplot +[Eta_rLimit] table[x=ka,y=General] {Data_RadEff_Curve4Insets.dat};
    \addplot +[Eta_rLimit_HighAlphaN] table[x=ka,y expr=\thisrow{Eta_r_sigma_highAlphaN}] {Data_RadEff_Curve4Insets.dat};

    \end{axis}
    \end{tikzpicture}%
}] at (pt2) {};

\end{tikzpicture}
	\caption{\textbf{Efficiency trend over different frequency ranges.} The left inset shows the efficiency at the $ka\leq 0.01$. The right inset illustrates efficiency over middle range $0.01\leq ka \leq 0.15$, where the frequency $f_i$ with the maximum slope is observed. 
	The limit is derived for a cylindrical dipole with total length and radius of \SI{151}{\milli\meter} and \SI{0.6745}{\milli\meter}, respectively.}
	\label{Fig_MaxSlope}
\end{figure}

One can rearrange \eqref{Eq_Eta_Bound_HighAlphaN} in the form $\eta_{r_{\max}} = \frac{bf\sqrt{f}}{bf\sqrt{f} + 1}$ where $b = \frac{4S}{{3c}^{2}}\sqrt{\frac{{\sigma \pi}}{\epsilon_{0}}}$ with \(c\) is the speed of light. 
The trend of radiation efficiency with increasing frequency is illustrated in Fig.~\ref{Fig_MaxSlope} over various intervals. 
The main graph shows $\eta_{r}$ in a broad frequency range, however, the small right inset shows $\eta_{r}$ in the vicinity of the point of maximum slope. 
The inset on the left shows the efficiency in the low frequency regime. 
It should be noted that efficiency has a form of \(f\sqrt{f}\) at low frequencies (left inset), however, after passing the point of maximum slope the rate of increase in efficiency becomes gradual. 
We can find the roll-over frequency from the second derivative of $\eta_{r}$. 
The roll-over point \(f_{i} = \left( 5b \right)^{- 2/3}\) which by substituting \(b\), we have:
\begin{align}
f_{i} = \sqrt[3]{\frac{{9\epsilon}_{0}c^{4}}{400\pi\sigma S^{2}}}
\end{align}
It is interesting to note that the efficiency has the fixed value of \(\frac{1}{6}\) at \(f_{i}\). 
This point can be used as a reference frequency for the transition between different regions: 
(a) the region with rapid changes in efficiency with frequency and (b) the region with the slower changes at higher efficiency levels.

\subsection{Variation of \(\eta_{r}\) with electrical area \(k^{2}S\)}
\label{Subsec_k2S}

\begin{figure}
	\centering
	\begin{tabular}{cp{10pt}c p{10pt} c}
	\includegraphics[height=4cm]{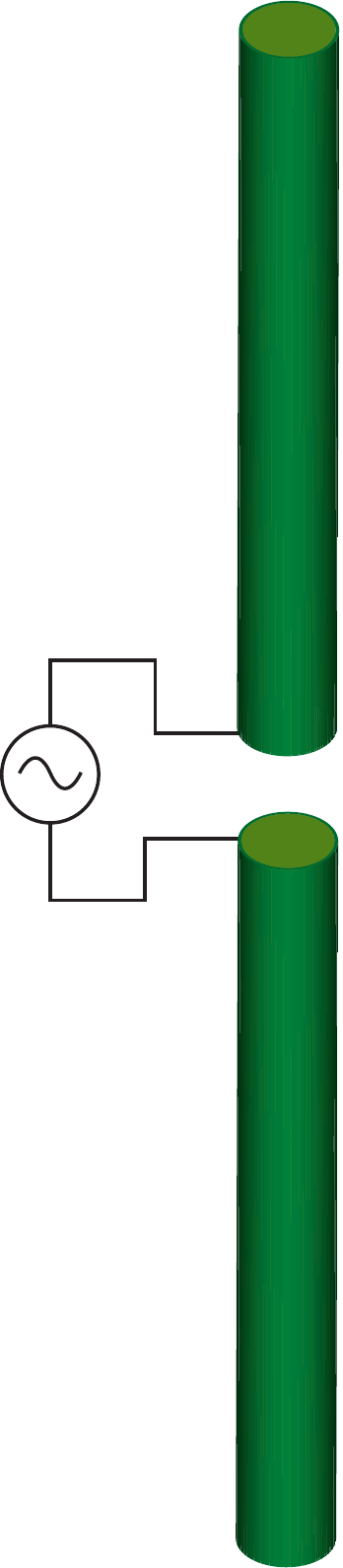}
	&
	&
	\includegraphics[height=4cm]{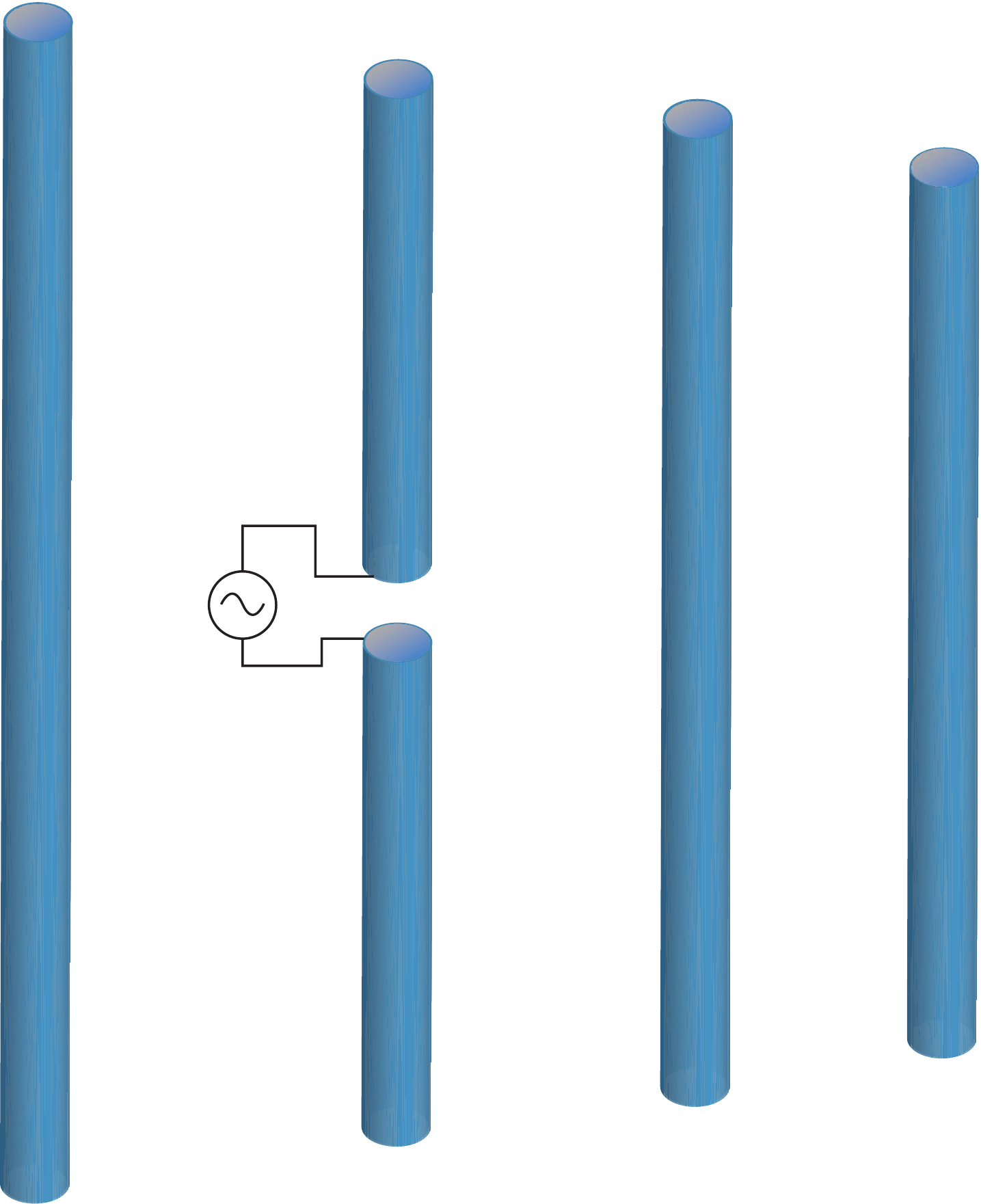}
	&
	&
	\includegraphics[height=2cm]{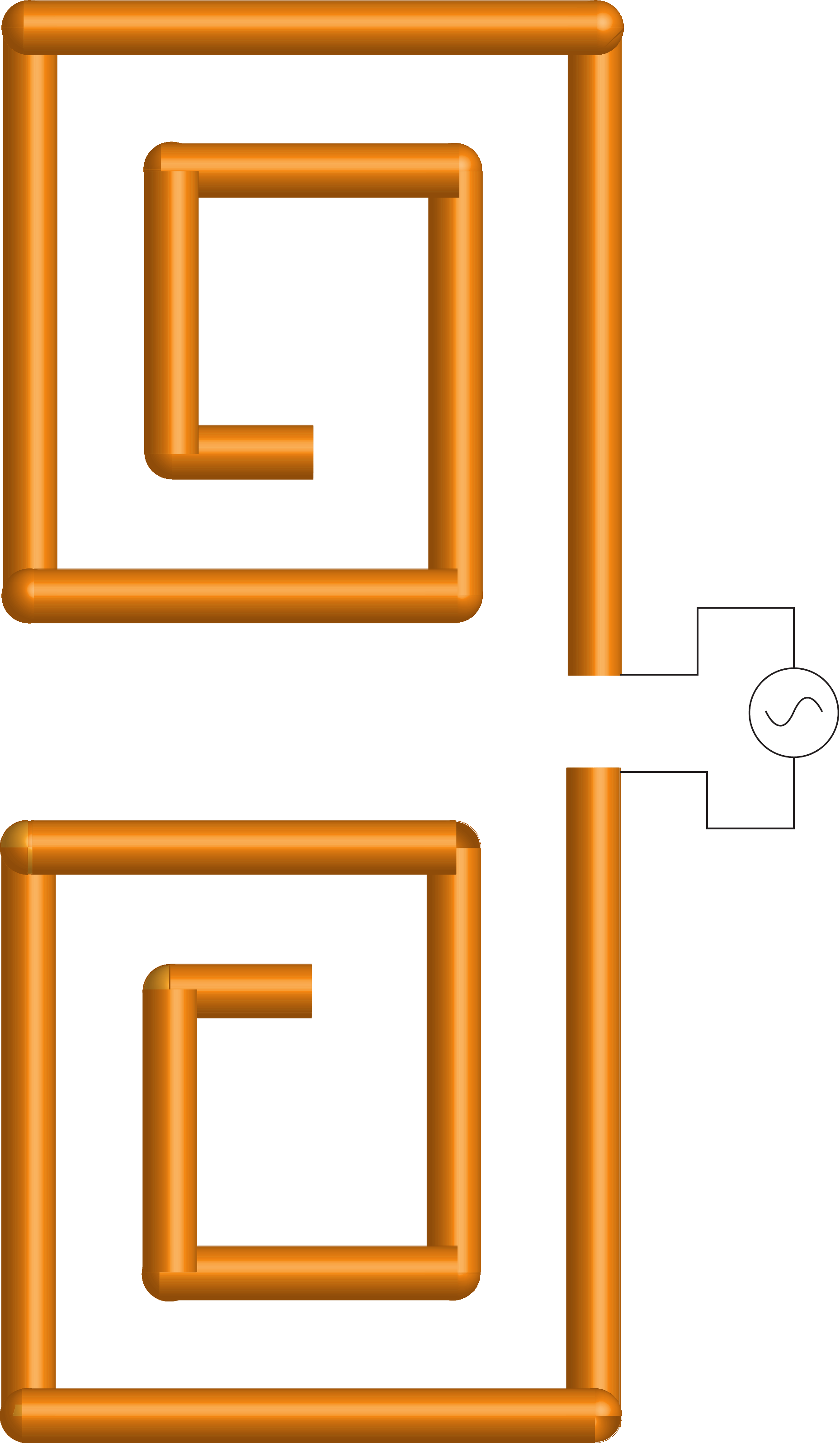}	\\
	(a) & & (b) & & (c)
	\end{tabular}	
	\caption{\textbf{Three different cylindrical wire antenna structures used for efficiency calculations}; (a) straight dipole, (b) Yagi-Uda and (c) meander line. 
	The wire radius was \SI{0.6745}{\milli\meter} and the resonant frequency was \SI{1}{\giga\hertz}.}
	\label{Fig_DipYM}
\end{figure}

Many studies~\cite{Chu_1948_JAP,Collin_1964_TAP,McLean_1996_TAP,Wheeler_1959_ProcIRE} reported the significance of the electrical length \(ka\), or even actual volume $V$~\cite{Vandenbosch_2012_TAP,Fujita_2015_IEICE} on the parameters like \(Q\) factor, gain, etc where \(a\) is the radius of the smallest sphere that encloses the whole antenna. 
Dipole, Yagi-Uda, and meander line antennas (see Fig.~\ref{Fig_DipYM}) were modeled using $\sigma =$ \SI{5.8E7}{\siemens\per\meter} and $n_{0} =$ \SI{0.6745}{\milli\meter}.
The surface area of the dipole, Yagi-Uda, and meanderline are \SI{5.9}{\centi\meter\squared}, \SI{23}{\centi\meter\squared}, and \SI{7.2}{\milli\meter\squared}, respectively.

The antennas are self-resonant almost at \SI{1}{\giga\hertz} while at other frequencies an ideal inductor is used to tune the antennas into resonance.
It should be noted that a realistic inductor can have significantly high Ohmic losses which further reduces the total efficiency but not radiation efficiency~\cite{Smith_1977_TAP}.
Smith~\cite{Smith_1977_TAP} provides a detailed analysis of the effect of the matching network loss on the total efficiency of the antenna.\footnote{Similar problem is partially addressed in~\cite{Jelinek_2017_arXiv}, but the authors mix the total efficiency and radiation efficiency.}

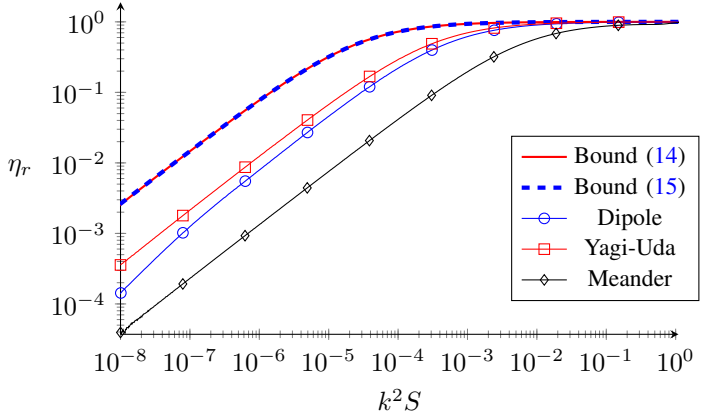
\begin{figure}
	\centering
		\pgfplotsset{
		AntennaDesign/.style={
			mark repeat = 50},
		Eta_rLimit/.style={
			mark = no marks, red,thick},
		Eta_rLimit_HighAlphaN/.style={
			mark = no marks, blue,dashed,ultra thick},
		Eta_Arbabi/.style={
			mark= no mark,teal ,ultra thick}
	}

\begin{tikzpicture}
\begin{loglogaxis}[xlabel=$k^2S$,ylabel=$\eta_r $, ylabel style={rotate={-90}},xmin=1e-8,xmax=1.1, ymax=1.9,,
width=9cm,height=6cm, axis y line= left, axis x line = bottom,]

\addplot [Eta_rLimit] table[x=ks2limit,y=eta_limit_G] {Data_Eta_vs_k_2_S_DYM.dat};

\addplot [Eta_rLimit_HighAlphaN] table[x=ks2limit,y=eta_limit_App] {Data_Eta_vs_k_2_S_DYM.dat};

\addplot [AntennaDesign,mark=o, blue] table[x=ks2Dip	,y=Eta_Dip] {Data_Eta_vs_k_2_S_DYM.dat};
\addplot [AntennaDesign,mark=square,red]table[x=ks2Yagi,y=Eta_Yagi] {Data_Eta_vs_k_2_S_DYM.dat};
\addplot [mark repeat=100, mark=diamond]table[x expr=\thisrow{ks2ML},y=Eta_ML] {Data_Eta_vs_k_2_S_DYM.dat};

\legend{\small Bound~\eqref{Eq_Eta_BoundGeneral}, \small Bound~\eqref{Eq_Eta_Bound_HighAlphaN}, \small Dipole,\small Yagi-Uda, \small Meander}

 	\pgfplotsset{every axis legend/.append style={
 			at={(0.7,0.1)},
 			anchor=south west}}
 			
\end{loglogaxis}
\end{tikzpicture}	
	\caption{\textbf{The new antenna efficiency $\eta_r$ on $k^2 S$ scale:} The three antennas are shown in Fig.~\ref{Fig_DipYM}. Equations \eqref{Eq_Eta_BoundGeneral} and \eqref{Eq_Eta_Bound_HighAlphaN} are the general and approximate bounds for structures with electrical area $k^2 S$.}
	\label{Fig_k2S}
\end{figure}

The bound from \eqref{Eq_Eta_BoundGeneral} is dependent on \(\sigma\), $\delta$  $n_{0}$, and \(k^{2}S\). 
Since the surface area of these antennas are different, their prospective upper bounds are not identical.
Mapping the antennas on the $k^2S$ scale is the only way to compare the performance of these antennas with fundamental limits in one graph (see Fig.~\ref{Fig_k2S}). 
This illustrates that different antennas have similar trends in efficiency when scaled on the \(k^{2}S\) axis. 
Therefore, we deduce that electrical area \(k^{2}S\) can be a valuable scale to compare the performance of different antennas. 
To the best of our knowledge, it is the first time that an investigation reveals the significance of the electrical area \(k^{2}S\) on the performance of the antenna.

\subsection{Variation of $\eta_r$ with conductivity}
\label{Subsec_Conductivity}

The consumer market highly demands conductive polymers, graphene and conductive inks for green and flexible electronics applications.
However, these novel materials often have low conductivities in comparison to copper. 
We reported an analysis of the influence of conductivity on efficiency, gain, cross sections, etc. in~\cite{Shahpari_2015_TAP}. 
It is  important to see how a reduction in conductivity can impact on antenna efficiency and its physical bounds.

\begin{figure}
	\centering
\newcommand{\D}{\mathrm{d}}
\newcommand{\FigWidth}{7.5cm}
\newcommand{\FigHeight}{5.5cm}

	\begin{tikzpicture}
	\pgfplotsset{every axis legend/.append style={
			at={(0.95,0.2)},
			anchor=south east}}

	\pgfplotsset{
	AntennaDesign/.style={
	mark = square,},
	Eta_rLimit/.style={
	mark = no marks, red,},
	Eta_rLimit_HighAlphaN/.style={
		mark = no marks, blue,dashed,thick},
	}

	\begin{groupplot}[width = \FigWidth, height = \FigHeight,xlabel=$\sigma$(\si{\siemens\per\meter}) , ylabel=$\eta_r$,
	y label style = {rotate={-90}},
    group style={group name={my plots},
     	group size=1 by 3,
     	 ylabels at= edge left,
 	     vertical sep=1.5cm
	     }, 
	     ymin=0, ymax=1,
	     xmode=log, 
	     title style={at={(-0.2,0.85)}}, 
	    ]

\nextgroupplot[title=\textbf{\Large (a)}]

\addplot [Eta_rLimit] table[x=sigma,y=Dipole] {DataRadEff_Limit.dat};
\addplot [Eta_rLimit_HighAlphaN] table[x=sigma,y=Dipole] {DataRadEff_Limit_ApproxHighAlphaN.dat};

\addplot [AntennaDesign] table[x=sigma,y=Dipole] {DataRadEff.dat};

\legend{\small Bound~\eqref{Eq_Eta_BoundGeneral}, \small Bound~\eqref{Eq_Eta_Bound_HighAlphaN}, \small Simulations}

\nextgroupplot[title=\textbf{\Large (b)}]

\addplot [Eta_rLimit] table[x=sigma,y=Yagi] {DataRadEff_Limit.dat};

\addplot [Eta_rLimit_HighAlphaN] table[x=sigma,y=Yagi] {DataRadEff_Limit_ApproxHighAlphaN.dat};

\addplot [AntennaDesign] table[x=sigma,y=Yagi] {DataRadEff.dat};

\legend{\small Bound~\eqref{Eq_Eta_BoundGeneral}, \small Bound~\eqref{Eq_Eta_Bound_HighAlphaN}, \small Simulations}

\nextgroupplot[title=\textbf{\Large (c)}]
\addplot [Eta_rLimit] table[x=sigma,y=Meander] {DataRadEff_Limit.dat};
\addplot [Eta_rLimit_HighAlphaN] table[x=sigma,y=Meander] {DataRadEff_Limit_ApproxHighAlphaN.dat};

\addplot [AntennaDesign] table[x=sigma,y=Meander] {DataRadEff.dat};

\legend{\small Bound~\eqref{Eq_Eta_BoundGeneral}, \small Bound~\eqref{Eq_Eta_Bound_HighAlphaN}, \small Simulations}

\end{groupplot}
\end{tikzpicture}
	\caption{\textbf{Variation of the efficiency $\eta_{r}$ with conductivity}: The three antennas were tuned to resonate at \SI{1}{\giga\hertz}. 
		Limitations from the general bound \eqref{Eq_Eta_BoundGeneral} and approximate formulas \eqref{Eq_Eta_Bound_HighAlphaN} compared to simulations for (a) dipole (b) Yagi-Uda and (c) meander line antennas.}
	\label{Fig_EtaLimit_eta_vs_sigma}
\end{figure}
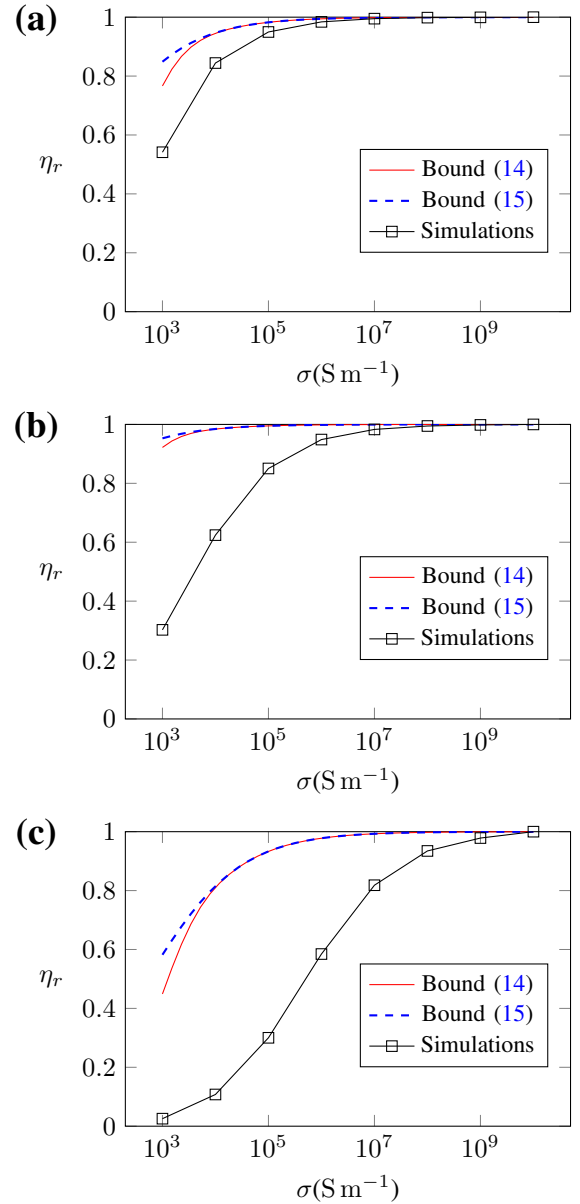

A comparison of the limitations proposed in this paper with different antennas are illustrated in Fig.~\ref{Fig_EtaLimit_eta_vs_sigma} for different values of conductivity \(\sigma\). 
The efficiency of a dipole, Yagi-Uda, and meander line antennas are compared with our general and approximate bounds. 
The antennas operate at \(f =\) \SI{1}{\giga\hertz}. 
It is seen from Fig.~\ref{Fig_EtaLimit_eta_vs_sigma} that both bounds are higher than the simulated value. 
It should be noted that each antenna has a different size, Chu radius \(a\), and occupies a different area \(S\). 
Therefore, the physical limitations of each individual antenna is different. 
For all three antennas, the approximate limit starts diverging from the general limit around \(\sigma \approx\) \SI{3000}{\siemens\per\meter} (which is almost 0.005\% of conductivity of copper). 
It should be noted that at this point we have: \(\alpha n_{0} \approx 3\) (see Fig.~\ref{Fig_alpha_n0}). 
Therefore, the necessary conditions for the approximations made in the derivation of \eqref{Eq_Eta_Bound_HighAlphaN} are not satisfied. 
This explains why the approximate formula cannot follow the general bound for the low conductive edge of the curve.
\subsection{Comparison with previous works}
\label{Subsec_CompareWOthers}
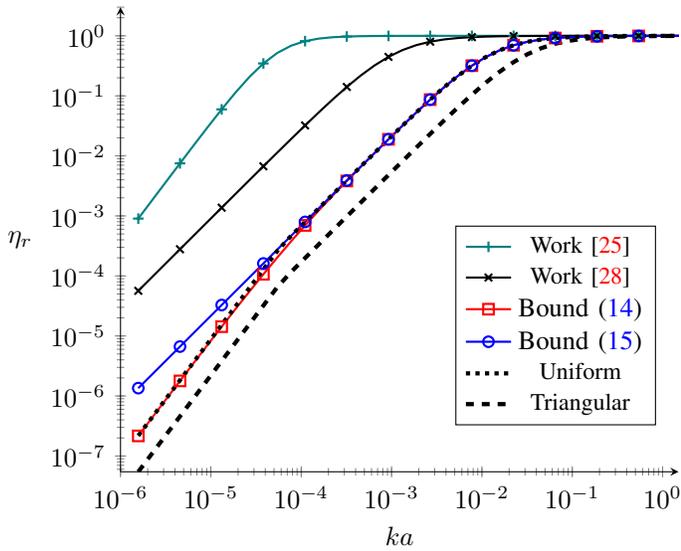
\begin{figure}
	\centering
		\pgfplotsset{
		AntennaDesign/.style={
			ultra thick,mark repeat = 3},
		Eta_rLimit/.style={
			mark =square,mark repeat = 3, red,thick},
		Eta_rLimit_HighAlphaN/.style={
			mark =o,mark repeat = 3, blue,thick},
		Eta_Arbabi/.style={
			mark=+, mark repeat = 3,teal ,thick},
		Eta_Pfeiffer/.style={
			mark=x,  mark repeat = 3,black ,thick}
	}

\begin{tikzpicture}

\begin{loglogaxis}[xlabel=$ka$,ylabel=$\eta_r$, y label style={rotate={-90}}, xmax=1.5, xmin=1e-6, ymax=3,
axis y line= left, axis x line = bottom,
width=9cm]

\addplot [Eta_Arbabi] table[x=ka,y=eta_Arbabi_Fujita1] {DataRadEff_ka_Sweep.dat};
	
\addplot [Eta_Pfeiffer] table[x=ka,y=Eta_Pfeiffer] {DataRadEff_ka_Sweep.dat};

\addplot [Eta_rLimit] table[x=ka,y=Eta_r_sigmad] {DataRadEff_ka_Sweep.dat};
	
\addplot [Eta_rLimit_HighAlphaN] table[x=ka,y=Eta_r_sigma_highAlphaN] {DataRadEff_ka_Sweep.dat};

\addplot [AntennaDesign, mark repeat = 2,dotted] table[x=ka,y=Eta_Analytic_inf] {DataRadEff_ka_Sweep.dat};

\addplot [AntennaDesign, mark repeat = 2,dashed] table[x=ka,y=Eta_Analytic_ShortSD] {DataRadEff_ka_Sweep.dat};
	
\legend{\small Work~\cite{Arbabi_2012_TAP}, \small Work \cite{Pfeiffer_2017_TAP}, Bound~\eqref{Eq_Eta_BoundGeneral}, Bound~\eqref{Eq_Eta_Bound_HighAlphaN}, \small Uniform, \small Triangular}

\pgfplotsset{every axis legend/.append style={
	at={(.6,.1)},
	anchor=south west}}
\end{loglogaxis}

\end{tikzpicture}
	\caption{\textbf{The efficiency variations for a straight wire dipole in terms of $ka$.} 
	Our new precise bound using \eqref{Eq_Eta_BoundGeneral}, the approximation \eqref{Eq_Eta_Bound_HighAlphaN} and the analytical solution for a small dipole, demonstrate that the previous efficiency bounds are greatly inflated compared to our new bounds.}
	\label{Fig_RadEff_ka_Sweep}
\end{figure}

We provide a comparison of the findings of the current paper with previously published bounds \cite{Arbabi_2012_TAP,Fujita_2015_IEICE,Pfeiffer_2017_TAP} and the analytic expected
values for small dipoles. 
Efficiency of a small dipole (with triangular current distribution) was computed from the work of Best and Yaghjian~\cite{Best_2004_AWPL_Lossy}.
Similarly, efficiency of a Hertzian dipole (with uniform distribution) was calculated which is almost four times higher than small dipole at \(ka \rightarrow 0\). 
The radiation efficiency of a {straight wire dipole} with length \(a =\) \SI{75}{\milli\meter} and radius \(r =\) \SI{0.675}{\milli\meter} was studied across the frequency range 100kHz to \SI{1}{\giga\hertz}.
The antenna conductivity was set to that of copper ($\sigma =$ \SI{5.6E7}{\siemens\per\meter}). 
A gap between bounds in~\cite{Arbabi_2012_TAP,Fujita_2015_IEICE} and the analytic solution for $\eta_{r}$ is evident in Fig.~\ref{Fig_RadEff_ka_Sweep} which widens as \(ka \rightarrow 0\). 
Maximum efficiency predicted from~\cite{Pfeiffer_2017_TAP} and also this work provided a tighter bound on maximum efficiency in comparison to~\cite{Arbabi_2012_TAP}.
The dissipation factor of~\cite{Pfeiffer_2017_TAP} for a short dipole $TM_{10}$ is given as $\frac{5\delta}{8ka^2}$ and it is still an order of magnitude higher than the theoretical values of a short dipole. 
The dissipation factor based on this work is $\frac{3\pi\delta}{2kS}$ which takes into account the actual area $S$ of cylindrical dipole rather Chu sphere.
That is, our new bound provides more accurate estimations of $\eta_{r}$ particularly at low \(ka\) values. 
Otherwise, by setting $S=4\pi a^2$, efficiency from this work would be close but slightly larger than  Pfeiffer~\cite{Pfeiffer_2017_TAP}.
We also see that a uniform current is the optimum distribution (in terms of radiation efficiency) for a dipole shaped radiator since it closely follows the physical bound on the efficiency.

We acknowledge that some researchers utilise optimisation algorithms to find the maximum possible efficiency of specific antenna shapes and fundamental limit~\cite{Jelinek_2017_TAP}. 
Numerically based optimization methods
\cite{Jelinek_2017_TAP}
have been shown to provide efficiency values which are very close to those calculated from our method for a specific antenna configuration and so very close to the fundamental limit. The difference between the optimization technique and the method presented here to obtain a fundamental limit lies in the formulation and solution of the efficiency calculation:
In a numerical optimisation approach the antenna is discretised and a discrete set of basis function are used to determine the current in each segment. From these values, radiated and lost powers are determined and so the efficiency is calculated.
In the approach presented here the antenna is not discretised and the current is described by a continuous function. 
The efficiency is determined by the finite conductivity of the materials in the antenna. 
The result is a formulation dependent on frequency and conductivity (without discretisation).

\begin{figure}
	\centering
\pgfplotsset{
	AntennaDesign/.style={
		black, thick,mark =diamond},
	Eta_rLimit/.style={
		mark = no marks, red,thick},
	Eta_rLimit_HighAlphaN/.style={
		mark = no marks, blue,dashed,ultra thick},
	Eta_Arbabi/.style={
		mark= no mark,teal ,thick}
}
\begin{tikzpicture}
\begin{semilogxaxis}[xlabel=Conductivity $\sigma$ (\si{\siemens\per\Box}),ylabel=Efficiency $\eta_r$,
axis y line= left, axis x line = bottom,
ymax=1.1,ymin=0.35,xmax=170,xmin=0.1]

\addplot [AntennaDesign, only marks] coordinates {((0.1,0.3761) (1,0.8590) (10,0.9829) (100,0.9957)};
\addplot [Eta_rLimit] table[x=sigma,y=MaxEff] {
	sigma				MaxEff
	0.100000000000000	0.400186634632703
	0.126896100316792	0.458473610278641
	0.161026202756094	0.517919669615297
	0.204335971785694	0.576862768069569
	0.259294379740467	0.633696128344648
	0.329034456231267	0.687037341836856
	0.417531893656040	0.735848844079061
	0.529831690628371	0.779490836511602
	0.672335753649934	0.817708690346207
	0.853167852417281	0.850572534319728
	1.08263673387405	0.878392731532290
	1.37382379588326	0.901632408511913
	1.74332882219999	0.920831164418103
	2.21221629107045	0.936546536190039
	2.80721620394118	0.949314048413483
	3.56224789026244	0.959623354961882
	4.52035365636024	0.967906681031976
	5.73615251044868	0.974535763995018
	7.27895384398315	0.979824109389411
	9.23670857187386	0.984032180361070
	11.7210229753348	0.987373886802069
	14.8735210729351	0.990023339387049
	18.8739182213510	0.992121269414121
	23.9502661998749	0.993780810069848
	30.3919538231320	0.995092521559137
	38.5662042116347	0.996128653049130
	48.9390091847749	0.996946694051563
	62.1016941891562	0.997592295431845
	78.8046281566991	0.998101648597591
	100	0.998503409063946
};
\legend{Optimisation~\cite{Gustafsson_2013_APS}, 2D Bound~\eqref{Eq_Bound2D}}
\pgfplotsset{every axis legend/.append style={
		at={(.5,.5)},
		anchor=south west}}
\end{semilogxaxis}
\end{tikzpicture}
	\caption{Comparison of the efficiency of the optimized antennas\cite{Gustafsson_2013_APS} versus the limitations proposed for 2D structures.}
	\label{Fig_EtaLimit_CompGust}
\end{figure}
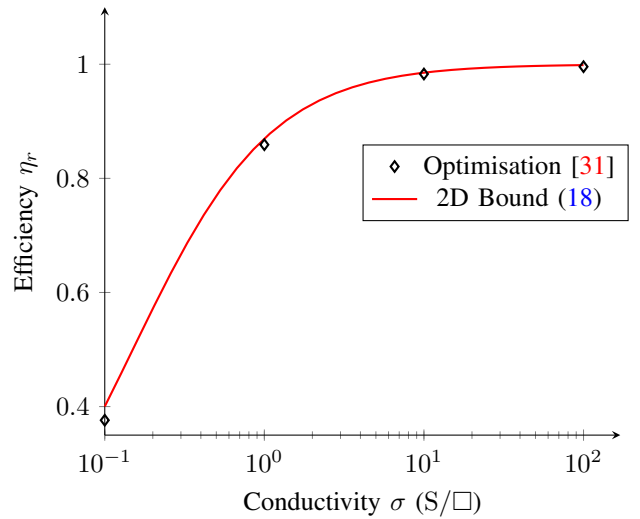

Figure~\ref{Fig_EtaLimit_CompGust} illustrates a comparison between the lossy planar antennas optimized~\cite{Gustafsson_2013_APS} using convex algorithm~\cite{Gustafsson_2016_FERMAT} with our 2D limit.
Efficiency of the optimized antennas are below but close to the maximum predicted efficiency. 
This can be considered as another validation of the presented approach.

\section{Conclusion}\label{conclusion}

In this article, we introduced new fundamental limits~\eqref{Eq_Eta_BoundGeneral} and \eqref{Eq_Bound2D} for the efficiency of the small antennas. 
The limit applies to antennas made from bulk homogenous materials, and also thin conductive sheets. 
Only three descriptors are needed in our efficiency calculations:
conductivity, frequency, and the antenna dimensions. 
This bound can predict the efficiency of the antennas more accurately than the previous contributions. 
Also, it can provide estimations for non-spherical antennas (e.g. planar structures).
The total electrical area \(k^{2}S\) has potential for future studies on antenna physical bounds. 
The impact of low \(\sigma\) on the $\eta_{r}$ limitations was explored and compared with simulated efficiency of the dipole, meander, and Yagi-Uda antennas. 
Based on the results of this paper, the maximum efficiency is achieved if the designer could distribute a uniform current over the antenna structure.


For many cases, the limit can be expressed as an approximation in a simple closed form \eqref{Eq_Eta_Bound_HighAlphaN}. 
The approximation is based on assuming the fields decrease exponentially from the surface and the Schwarz inequality. 
Simple approximations enable the calculation of new upper bounds on the radiation efficiency $\eta_{r}$ for frequencies much less than the plasma frequency of the conductor. 
In the case of a lossy metallic structure where the conductor thickness is much larger than the skin depth, this approximate formula gives accurate results. 
At low frequencies, the efficiency increases with \(f^{1.5}\) factor.
At high frequencies the efficiency curve is in the form \(\frac{bf^{1.5}}{bf^{1.5} + 1}\) (where $b$ is a constant).
The roll-over point in the curve depends on the conductivity and the total surface area.

The results and conclusions presented in this paper are particularly important as researchers investigate the use of laser induced conductive
polymers and graphene as conductive antenna elements. 
It can also provide the basis for the first fundamental limit on the efficiency of
an optical nanoantenna~\cite{Novotny_2011_nphot}, if the calculations here are properly
modified by the surface plasmon effect.

\appendices
\section*{Appendix}\label{appendix}

In this appendix a proof for the identity~\eqref{Eq_R2Div} is presented:
\begin{multline}
I= \int_{V_1} \int_{V_2} R^2\, \nabla_1 \cdot\boldsymbol{J}(\boldsymbol{r}_1) \, \nabla_2 \cdot\boldsymbol{J}^*(\boldsymbol{r}_2) \, \mathrm{d}V_1 \, \mathrm{d}V_2 \\
=-2\int_{V_1} \int_{V_2}  \boldsymbol{J}(\boldsymbol{r}_1) \cdot \boldsymbol{J}^*(\boldsymbol{r}_2) \, \mathrm{d}V_1 \, \mathrm{d}V_2
\label{Eq_AppIdentity}
\end{multline}
where $R=|\boldsymbol{r}_1-\boldsymbol{r}_2|$ and $\boldsymbol{r}_1$ and $\boldsymbol{r}_2$ are the position vectors on the volume $V_1$ and $V_2$, respectively.
Here, we refer to $\boldsymbol{J}(\boldsymbol{r}_1)$ and $\boldsymbol{J}(\boldsymbol{r}_2)$ by $\boldsymbol{J}_1$ and $\boldsymbol{J}_2$ for the sake of the simplicity of the notation.

Using the Green first identity~\cite{Arfken_b_2001,Kellogg_b_1929}, we have:
\begin{align}
\int_{V_1} \left|\boldsymbol{r}_1-\boldsymbol{r}_2\right|^2 &  \nabla_1 \cdot\boldsymbol{J}_1 \, \mathrm{d}V_1\nonumber\\
=& -\int_{V_1} \boldsymbol{J}_1 \cdot \nabla_1 \left|\boldsymbol{r}_1-\boldsymbol{r}_2\right|^2 \, \mathrm{d}V_1\nonumber \\
&+ \oint_{S_1} \left|\boldsymbol{r}_1-\boldsymbol{r}_2\right|  \boldsymbol{J}_1 \cdot \hat{\boldsymbol{n}} \, \mathrm{d}S_1\label{Eq_AppWIntS}\\
=& -2 \int_{V_1} \boldsymbol{J}_1 \cdot (\boldsymbol{r}_1-\boldsymbol{r}_2) \, \mathrm{d}V_1
\end{align}
The surface integral over $S_1$ in \eqref{Eq_AppWIntS} is omitted since current only flows on the surface $\boldsymbol{J}_1\cdot\hat{\boldsymbol{n}}=0$.

We start by LHS of \eqref{Eq_AppIdentity}:
\begin{align}
I =  -2\int_{V_1} \boldsymbol{J}_1\cdot  \int_{V_2} (\boldsymbol{r}_1-\boldsymbol{r}_2) \nabla_2\cdot\boldsymbol{J}^*_2 \, \mathrm{d}V_2 \, \mathrm{d}V_1
\label{Eq_AppDropr1}
\end{align}
One can drop $\boldsymbol{r}_1$ in \eqref{Eq_AppDropr1} since it is multiplied by $\int \nabla_2\cdot\boldsymbol{J}_2\, \mathrm{d}V_2$.
By using the Green's first identity one more time, we have:
\begin{align}
\label{Eq_AppGIusedagain}
I =  -2\int_{V_1} \int_{V_2} \boldsymbol{J}_2^*\cdot \nabla_2\left[\boldsymbol{J}_1 \cdot \boldsymbol{r}_2\right] \, \mathrm{d}V_2 \, \mathrm{d}V_1.
\end{align}

Using the gradient of the dot product identity $\nabla[\boldsymbol{A}\cdot\boldsymbol{B}] =(\boldsymbol{A}\cdot\nabla)\boldsymbol{B} + (\boldsymbol{B}\cdot\nabla)\boldsymbol{A} +\boldsymbol{A}\times(\nabla\times\boldsymbol{B}) +\boldsymbol{B}\times(\nabla\times\boldsymbol{A})$, we write:
\begin{align}
\label{Eq_AppDelDotJ}
\nabla_2 \left[\boldsymbol{J}_1 \cdot \boldsymbol{r}_2 \right]
= \boldsymbol{J}_1.
\end{align}
Using~\eqref{Eq_AppDelDotJ} in \eqref{Eq_AppGIusedagain}, we get:
\begin{align}
I = -2 \int_{V_1} \int_{V_2} \boldsymbol{J}^*_2 \cdot \boldsymbol{J}_1 \, \mathrm{d}V_1 \, \mathrm{d}V_2
\end{align}

Since the dot product is a commutative operator $\boldsymbol{J}^*_2 \cdot \boldsymbol{J}_1 = \boldsymbol{J}_1 \cdot \boldsymbol{J}^*_2$. 
Therefore, we have $I$ in the exact form of RHS of \eqref{Eq_AppIdentity}.
This concludes the proof.


\newpage

\begin{IEEEbiography}[{\includegraphics[width=1in,height=1.25in,clip,keepaspectratio]{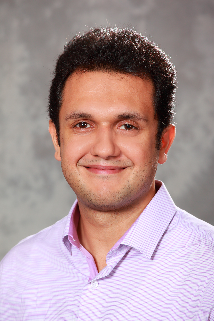}}]{Morteza Shahpari}
\textbf{(SM-08,M15)}
 received the Bachelor’s and Master’s degrees in telecommunications engineering from Iran University of Science and Technology (IUST), Tehran, Iran, in 2005 and 2008, respectively. He received his Ph.D. from Griffith University, Brisbane, Australia in 2015 where he focused on the fundamental limitations of small antennas. He is currently teaching as an associate lecturer in Griffith University.

His research interests include fundamental limitations of antennas, equivalent circuits for antennas, and antenna scattering as well as carbon nanotube and graphene antennas. He is also interested in mathematical methods in electromagnetics particularly Green's functions.
\end{IEEEbiography}

\vspace*{-4\baselineskip}
\begin{IEEEbiography}[{\includegraphics[width=1in,height=1.25in,clip,keepaspectratio]{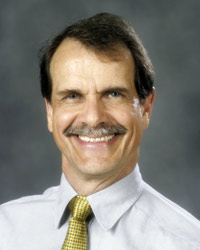}}]
{David V. Thiel}
\textbf{(M’81–SM’88–LSM’16)} received the bachelor’s degree in physics and applied mathematics from the University of Adelaide, Adelaide, SA, Australia, and the M.S. and Ph.D. degrees from James Cook University, Townsville, QLD., Australia.He is currently the Deputy Head of the Griffith School of Engineering, Griffith University, Brisbane, QLD, Australia. He has authored the book Research Methods for Engineers (Cambridge, U.K.: Cambridge University Press, 2014), and co-authored a book on Switched Parasitic Antennas for Cellular Communications (Norwood, MA, USA: Artech House, 2002). He has authored six book chapters, over 140 journal papers, and co-authored more than nine patent applications. His current research interests include electromagnetic geophysics, sensor development, electronics systems design and manufacture, antenna development for wireless sensor networks, environmental sustainability in electronics manufacturing, sports engineering, and mining engineering.Prof. Thiel is a fellow of the Institution of Engineers, Australia, and a Chartered Professional Engineer in Australia.
\end{IEEEbiography}

\vfil


\begin{thebibliography}{10}
	\providecommand{\url}[1]{#1}
	\csname url@samestyle\endcsname
	\providecommand{\newblock}{\relax}
	\providecommand{\bibinfo}[2]{#2}
	\providecommand{\BIBentrySTDinterwordspacing}{\spaceskip=0pt\relax}
	\providecommand{\BIBentryALTinterwordstretchfactor}{4}
	\providecommand{\BIBentryALTinterwordspacing}{\spaceskip=\fontdimen2\font plus
		\BIBentryALTinterwordstretchfactor\fontdimen3\font minus
		\fontdimen4\font\relax}
	\providecommand{\BIBforeignlanguage}[2]{{%
			\expandafter\ifx\csname l@#1\endcsname\relax
			\typeout{** WARNING: IEEEtran.bst: No hyphenation pattern has been}%
			\typeout{** loaded for the language `#1'. Using the pattern for}%
			\typeout{** the default language instead.}%
			\else
			\language=\csname l@#1\endcsname
			\fi
			#2}}
	\providecommand{\BIBdecl}{\relax}
	\BIBdecl
	
	\bibitem{AntDef_2014_IEEESTD}
	``{IEEE Standard for Definitions of Terms for Antennas},'' \emph{IEEE Std
		145-2013 (Revision of IEEE Std 145-1993)}, pp. 1--50, mar 2014.
	
	\bibitem{Volakis_b_2010}
	J.~L. Volakis, C.~C. Chen, and K.~Fujimoto, \emph{{Small antennas
			miniaturization techniques and applications}}.\hskip 1em plus 0.5em minus
	0.4em\relax McGraw-Hill, 2010.
	
	\bibitem{Gustafsson_2015_SpringerChapter}
	\BIBentryALTinterwordspacing
	M.~Gustafsson, D.~Tayli, and M.~Cismasu, ``{Physical Bounds of Antennas},'' in
	\emph{Handbook of Antenna Technologies}.\hskip 1em plus 0.5em minus
	0.4em\relax Singapore: Springer Singapore, 2015, pp. 1--32.
	\BIBentrySTDinterwordspacing
	
	\bibitem{Shahpari_2015_Thesis}
	M.~Shahpari, ``{Fundamental limitations of the small antennas},'' Griffith
	University, Ph.D. Thesis, 2015.
	
	\bibitem{Shahpari_2015_TAP}
	\BIBentryALTinterwordspacing
	M.~Shahpari and D.~V. Thiel, ``{The Impact of Reduced Conductivity on the
		Performance of Wire Antennas},'' \emph{IEEE Transactions on Antennas and
		Propagation}, vol.~63, no.~11, pp. 4686--4692, nov 2015.
	\BIBentrySTDinterwordspacing
	
	\bibitem{Harrington_1960_JRNBS}
	\BIBentryALTinterwordspacing
	R.~F. Harrington, ``{Effect of antenna size on gain, bandwidth, and
		efficiency},'' \emph{J. Res. Nat. Bur. Stand}, vol. 64D, no.~1, p.~1, 1960.
	\BIBentrySTDinterwordspacing
	
	\bibitem{Hansen_1981_PIEEE}
	R.~C. Hansen, ``{Fundamental limitations in antennas},'' pp. 170--182, 1981.
	
	\bibitem{Geyi_2003_TAP_PhysLim}
	\BIBentryALTinterwordspacing
	W.~Geyi, ``{Physical limitations of antenna},'' \emph{IEEE Transactions on
		Antennas and Propagation}, vol.~51, no.~8, pp. 2116--2123, aug 2003.
	\BIBentrySTDinterwordspacing
	
	\bibitem{Pigeon_2014_IJMWT}
	\BIBentryALTinterwordspacing
	M.~Pigeon, C.~Delaveaud, L.~Rudant, and K.~Belmkaddem, ``{Miniature directive
		antennas},'' \emph{International Journal of Microwave and Wireless
		Technologies}, vol.~6, no.~01, pp. 45--50, feb 2014.
	\BIBentrySTDinterwordspacing
	
	\bibitem{Wheeler_1947_ProcIEEE}
	\BIBentryALTinterwordspacing
	H.~Wheeler, ``{Fundamental limitations of small antennas},'' \emph{Proceedings
		of the IRE}, vol.~35, no.~12, pp. 1479--1484, dec 1947.
	\BIBentrySTDinterwordspacing
	
	\bibitem{Chu_1948_JAP}
	\BIBentryALTinterwordspacing
	L.~J. Chu, ``{Physical limitations of omni-directional antennas},''
	\emph{Journal of Applied Physics}, vol.~19, no.~12, p. 1163, 1948.
	\BIBentrySTDinterwordspacing
	
	\bibitem{Collin_1964_TAP}
	\BIBentryALTinterwordspacing
	R.~E. Collin and S.~Rothschild, ``{Evaluation of antenna Q},'' \emph{IEEE
		Transactions on Antennas and Propagation}, vol.~12, no.~1, pp. 23--27, jan
	1964.
	\BIBentrySTDinterwordspacing
	
	\bibitem{Fante_1969_TAP}
	\BIBentryALTinterwordspacing
	R.~Fante, ``{Quality factor of general ideal antennas},'' \emph{IEEE
		Transactions on Antennas and Propagation}, vol.~17, no.~2, pp. 151--155, mar
	1969.
	\BIBentrySTDinterwordspacing
	
	\bibitem{McLean_1996_TAP}
	\BIBentryALTinterwordspacing
	J.~S. McLean, ``{A re-examination of the fundamental limits on the radiation Q
		of electrically small antennas},'' \emph{IEEE Transactions on Antennas and
		Propagation}, vol.~44, no.~5, p. 672, may 1996.
	\BIBentrySTDinterwordspacing
	
	\bibitem{Thal_2006_TAP}
	\BIBentryALTinterwordspacing
	H.~L. Thal, ``{New radiation Q limits for spherical wire antennas},''
	\emph{IEEE Transactions on Antennas and Propagation}, vol.~54, no.~10, pp.
	2757--2763, oct 2006.
	\BIBentrySTDinterwordspacing
	
	\bibitem{Thal_2012_TAP}
	\BIBentryALTinterwordspacing
	------, ``{Q bounds for arbitrary small antennas: a circuit approach},''
	\emph{IEEE Transactions on Antennas and Propagation}, vol.~60, no.~7, pp.
	3120--3128, jul 2012.
	\BIBentrySTDinterwordspacing
	
	\bibitem{Kim_2016_TAP}
	\BIBentryALTinterwordspacing
	O.~S. Kim, ``{Lower Bounds on Q for Finite Size Antennas of Arbitrary Shape},''
	\emph{IEEE Transactions on Antennas and Propagation}, vol.~64, no.~1, pp.
	146--154, jan 2016.
	\BIBentrySTDinterwordspacing
	
	\bibitem{Jonsson_2015_RSPA}
	\BIBentryALTinterwordspacing
	B.~L.~G. Jonsson and M.~Gustafsson, ``{Stored energies in electric and magnetic
		current densities for small antennas},'' \emph{Proceedings of the Royal
		Society A: Mathematical, Physical and Engineering Sciences}, vol. 471, no.
	2176, pp. 20\,140\,897--20\,140\,897, mar 2015.
	\BIBentrySTDinterwordspacing
	
	\bibitem{Hansen_2011_APM}
	\BIBentryALTinterwordspacing
	P.~Hansen and R.~Adams, ``{The minimum Q for spheroidally shaped objects:
		extension to cylindrically shaped objects and comparison to practical
		antennas},'' \emph{IEEE Antennas and Propagation Magazine}, vol.~53, no.~3,
	pp. 75--83, jun 2011.
	\BIBentrySTDinterwordspacing
	
	\bibitem{Yaghjian_2013_PIER}
	A.~D. Yaghjian, M.~Gustafsson, and B.~L.~G. Jonsson, ``{Minimum {\{}Q{\}} for
		lossy and lossless electrically small dipole antennas},'' \emph{Progress In
		Electromagnetics Research}, vol. 143, pp. 641--673, 2013.
	
	\bibitem{Yaghjian_2005_TAP}
	\BIBentryALTinterwordspacing
	A.~D. Yaghjian and S.~R. Best, ``{Impedance, bandwidth, and Q of antennas},''
	\emph{IEEE Transactions on Antennas and Propagation}, vol.~53, no.~4, pp.
	1298--1324, apr 2005.
	\BIBentrySTDinterwordspacing
	
	\bibitem{Gustafsson_2007_RSPA}
	\BIBentryALTinterwordspacing
	M.~Gustafsson, C.~Sohl, and G.~Kristensson, ``{Physical limitations on antennas
		of arbitrary shape},'' \emph{Proceedings of the Royal Society A:
		Mathematical, Physical and Engineering Sciences}, vol. 463, no. 2086, pp.
	2589--2607, oct 2007.
	\BIBentrySTDinterwordspacing
	
	\bibitem{Gustafsson_2009_TAP}
	\BIBentryALTinterwordspacing
	------, ``{Illustrations of new physical bounds on linearly polarized
		antennas},'' \emph{IEEE Transactions on Antennas and Propagation}, vol.~57,
	no.~5, pp. 1319--1327, may 2009.
	\BIBentrySTDinterwordspacing
	
	\bibitem{Lewis_bc_2009}
	\BIBentryALTinterwordspacing
	A.~Lewis, M.~Randall, A.~Galehdar, D.~Thiel, and G.~Weis, ``{Using Ant Colony
		Optimisation to Construct Meander-Line RFID Antennas},'' in
	\emph{Biologically-Inspired Optimisation Methods}, ser. Studies in
	Computational Intelligence, A.~Lewis, S.~Mostaghim, and M.~Randall,
	Eds.\hskip 1em plus 0.5em minus 0.4em\relax Springer Berlin Heidelberg, 2009,
	vol. 210, pp. 189--217.
	\BIBentrySTDinterwordspacing
	
	\bibitem{Arbabi_2012_TAP}
	\BIBentryALTinterwordspacing
	A.~Arbabi and S.~Safavi-Naeini, ``{Maximum gain of a lossy antenna},''
	\emph{IEEE Transactions on Antennas and Propagation}, vol.~60, no.~1, pp.
	2--7, jan 2012.
	\BIBentrySTDinterwordspacing
	
	\bibitem{Fujita_2015_IEICE}
	K.~Fujita and H.~Shirai, ``{Theoretical limitation of the radiation efficiency
		for homogenous electrically small antennas},'' \emph{IEICE Transactions on
		Electronics}, vol. E98.C, no.~1, pp. 1--7, 2015.
	
	\bibitem{Karlsson_2013_PIER}
	\BIBentryALTinterwordspacing
	A.~Karlsson, ``\BIBforeignlanguage{en}{{On the efficiency and gain of
			antennas}},'' \emph{\BIBforeignlanguage{en}{Progress In Electromagnetics
			Research}}, vol. 136, pp. 479--494, 2013.
	\BIBentrySTDinterwordspacing
	
	\bibitem{Pfeiffer_2017_TAP}
	\BIBentryALTinterwordspacing
	C.~Pfeiffer, ``{Fundamental Efficiency Limits for Small Metallic Antennas},''
	\emph{IEEE Transactions on Antennas and Propagation}, vol.~65, no.~4, pp.
	1642--1650, apr 2017.
	\BIBentrySTDinterwordspacing
	
	\bibitem{Thal_2018_TAP}
	\BIBentryALTinterwordspacing
	H.~L. Thal, ``{Radiation efficiency limits for elementary antenna shapes},''
	\emph{IEEE Transactions on Antennas and Propagation}, pp. 1--1, 2018.
	\BIBentrySTDinterwordspacing
	
	\bibitem{Thal_1978_TAP}
	\BIBentryALTinterwordspacing
	------, ``{Exact circuit analysis of spherical waves},'' \emph{IEEE
		Transactions on Antennas and Propagation}, vol.~26, no.~2, pp. 282--287, mar
	1978.
	\BIBentrySTDinterwordspacing
	
	\bibitem{Gustafsson_2013_APS}
	\BIBentryALTinterwordspacing
	M.~Gustafsson, ``{Efficiency and Q for small antennas using Pareto
		optimality},'' in \emph{2013 IEEE Antennas and Propagation Society
		International Symposium (APSURSI)}.\hskip 1em plus 0.5em minus 0.4em\relax
	IEEE, jul 2013, pp. 2203--2204.
	\BIBentrySTDinterwordspacing
	
	\bibitem{Maier_b_2007}
	S.~A. Maier, \emph{{Plasmonics: Fundamentals and Applications}}.\hskip 1em plus
	0.5em minus 0.4em\relax Springer, 2007.
	
	\bibitem{Vandenbosch_2010_TAP}
	\BIBentryALTinterwordspacing
	G.~Vandenbosch, ``{Reactive energies, impedance, and Q factor of radiating
		structures},'' \emph{IEEE Transactions on Antennas and Propagation}, vol.~58,
	no.~4, pp. 1112--1127, apr 2010.
	\BIBentrySTDinterwordspacing
	
	\bibitem{Elliott_b_1982}
	\BIBentryALTinterwordspacing
	R.~S. Elliott, \emph{{Antenna theory and design}}.\hskip 1em plus 0.5em minus
	0.4em\relax IEEE, 2003.
	\BIBentrySTDinterwordspacing
	
	\bibitem{Balanis_2005_antenna}
	C.~A. Balanis, \emph{{Antenna theory: analysis and design}}, 3rd~ed.\hskip 1em
	plus 0.5em minus 0.4em\relax Wiley-Interscience, 2005.
	
	\bibitem{Galehdar_2007_AWPL}
	\BIBentryALTinterwordspacing
	A.~Galehdar, D.~Thiel, and S.~O'Keefe, ``{Antenna efficiency calculations for
		electrically small, RFID antennas},'' \emph{IEEE Antennas and Wireless
		Propagation Letters}, vol.~6, no.~11, pp. 156--159, 2007.
	\BIBentrySTDinterwordspacing
	
	\bibitem{Wheeler_1959_ProcIRE}
	\BIBentryALTinterwordspacing
	H.~Wheeler, ``{The radiansphere around a small antenna},'' \emph{Proceedings of
		the IRE}, vol.~47, no.~8, pp. 1325--1331, aug 1959.
	\BIBentrySTDinterwordspacing
	
	\bibitem{Vandenbosch_2012_TAP}
	\BIBentryALTinterwordspacing
	G.~A.~E. Vandenbosch, ``{Explicit relation between volume and lower bound for Q
		for small dipole topologies},'' \emph{IEEE Transactions on Antennas and
		Propagation}, vol.~60, no.~2, pp. 1147--1152, feb 2012.
	\BIBentrySTDinterwordspacing
	
	\bibitem{Smith_1977_TAP}
	\BIBentryALTinterwordspacing
	G.~Smith, ``{Efficiency of electrically small antennas combined with matching
		networks},'' \emph{IEEE Transactions on Antennas and Propagation}, vol.~25,
	no.~3, pp. 369--373, may 1977.
	\BIBentrySTDinterwordspacing
	
	\bibitem{Jelinek_2017_arXiv}
	\BIBentryALTinterwordspacing
	L.~Jelinek, K.~Schab, and M.~Capek, ``{The Radiation Efficiency Cost of
		Resonance Tuning},'' nov 2017.
	\BIBentrySTDinterwordspacing
	
	\bibitem{Best_2004_AWPL_Lossy}
	S.~R. Best and A.~D. Yaghjian, ``{The lower bounds on Q for lossy electric and
		magnetic dipole antennas},'' \emph{IEEE Antennas and Wireless Propagation
		Letters}, vol.~3, pp. 314--316, 2004.
	
	\bibitem{Jelinek_2017_TAP}
	\BIBentryALTinterwordspacing
	L.~Jelinek and M.~Capek, ``{Optimal Currents on Arbitrarily Shaped Surfaces},''
	\emph{IEEE Transactions on Antennas and Propagation}, vol.~65, no.~1, pp.
	329--341, jan 2017.
	\BIBentrySTDinterwordspacing
	
	\bibitem{Gustafsson_2016_FERMAT}
	M.~Gustafsson, D.~Tayli, C.~Ehrenborg, M.~Cismasu, and S.~Nordebo, ``{Antenna
		current optimization using MATLAB and CVX},'' \emph{FERMAT}, vol.~15, 2016.
	
	\bibitem{Novotny_2011_nphot}
	\BIBentryALTinterwordspacing
	L.~Novotny and N.~van Hulst, ``{Antennas for light},'' \emph{Nature Photonics},
	vol.~5, no.~2, pp. 83--90, feb 2011.
	\BIBentrySTDinterwordspacing
	
	\bibitem{Arfken_b_2001}
	G.~B. Arfken and H.~J. Weber, \emph{{Mathematical methods for physicists}},
	5th~ed.\hskip 1em plus 0.5em minus 0.4em\relax London: Academic press, 2001.
	
	\bibitem{Kellogg_b_1929}
	\BIBentryALTinterwordspacing
	O.~D. Kellogg, \emph{{Foundations of Potential Theory}}.\hskip 1em plus 0.5em
	minus 0.4em\relax Berlin, Heidelberg: Springer Berlin Heidelberg, 1929.
	\BIBentrySTDinterwordspacing
	
\end{thebibliography}
\end{document}